\documentclass[10pt,letterpaper,english]{article}

\usepackage{graphicx}
\usepackage{newtxtext}
\usepackage{newtxmath}
\usepackage{natbib}
\usepackage{hyperref}
\usepackage{enumitem}
\advance \topmargin by -\headheight
\advance \topmargin by -\headsep

\evensidemargin \oddsidemargin
\hypersetup{
    colorlinks = true,
    urlcolor   = blue,
    citecolor  = black,
}

\newcommand{\RomanNumeralCaps}[1]
\linenumbers

\title{Minimization Principle for Analytical Solution of Turbulent Flow in Channel}

\author{Alex Fedoseyev \\
email: \href{mailto:af@ultraquantum.com}{af@ultraquantum.com}
}
\date{%
Ultra Quantum Inc., \\
Huntsville, Alabama, USA\\
\today
}

\begin{document}
\maketitle

\begin{abstract}
The analytical solution for turbulent flow in channel presented in  \citet{Fedoseyev_2023},  described the mean turbulent flow velocity as a superposition of the laminar (parabolic) and turbulent (superexponential) solutions.  
In this study, the coefficients of superposition are proposed to obtain through the minimization principle, the principle of minimum viscous dissipation. The obtained analytical solutions agree well with the experimental data for turbulent flow.
\end{abstract}

%
%
\section{Introduction\label{sec:intro}}

An approximate analytical solution for turbulent flow in the channel was obtained in \citet{Fedoseyev_2023} by solving the Generalized Hydrodynamic Equations (GHE) proposed by \citet{Alexeev_1994}.
The solution for turbulent flow in a channel presented a mean turbulent flow velocity $U_{GHE}$ as a superposition of the laminar (parabolic) $U_L$ and turbulent (superexponential) $U_T$ solutions,
\begin{eqnarray}\label{eq:GHE1sol}
U_{GHE}=\gamma U_{T}+(1-\gamma)U_{L}, 
\end{eqnarray}

\noindent where the coefficients $\gamma$ and $(1-\gamma)$ were introduced.
The expressions for $U_T$ and $U_L$ were explicitly provided  giving 
\begin{eqnarray}\label{eq:GHE2sol}
U_{GHE}=U_{0}\left[\gamma\left(1-e^{1-e^{y/\delta}}\right)+(1-\gamma)4y(L-y)/L^{2}\right] 
\end{eqnarray}

\noindent in 2D channel, where $x$ is the coordinate along a channel, $y$ is the transversal coordinate, $L$ is the width of a channel with a centerline velocity $U_0$. All parameters are nondimensional. The parameters $\delta$ is  
\begin{eqnarray}\label{eq:delta}
\delta = {\sqrt{\tau^{*}\nu}}/{L_0},
\end{eqnarray}

\noindent where  $\tau^*$ is the relaxation time, or timescale, a material property for particular liquid or gas  used in the experiments,  $\nu$ is the kinematic viscosity, and $L_0$ is the hydrodynamic scale. The nondimensional $\tau$, a timescale coefficient for the fluctuation terms in GHE, is expressed as
\begin{equation}\label{eq:tau}
\tau = \tau^* L_0^{-1}U_{0}= \delta^2 Re,
\end{equation}
\noindent
where ${Re=U_0 L_0/\nu}$ denotes the Reynolds number. The analytical solution to Eq. \eqref{eq:GHE2sol} can also be used for the turbulent flow in circular pipe, if $\delta \ll 1$.

\begin{figure}
\begin{center}
\includegraphics[width=0.75\textwidth]{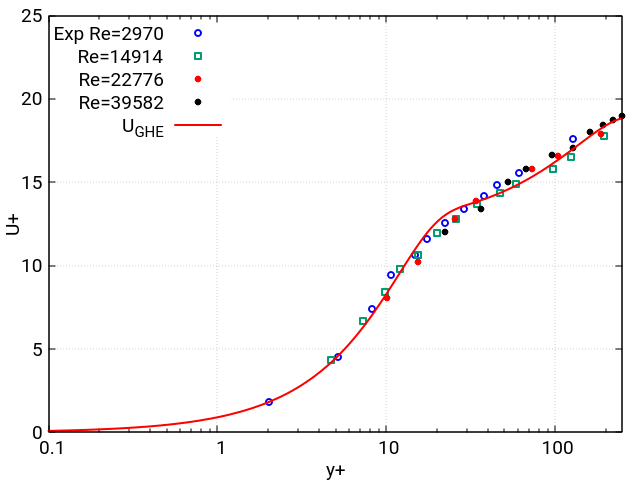}
\end{center}
\caption{\label{fig:exp0}Mean velocity profiles in turbulent boundary layer from \cite{Wei_1989} experiment (distilled water), non-dimensionalized on inner variables,
for the four Reynolds numbers: 2970 (circles); 14914
(squares); 22776 (red dots) and 39582 (black dots). The red line is an analytical GHE solution.}
\end{figure}

As an example, the analytical solution $U_{GHE}$ (red line) for the experiments of \cite{Wei_1989} is shown in Figure \ref{fig:exp0} in $(U^+, y^+)$ coordinates. The experimental velocity is shown as points for four Reynolds numbers. 
The parameter $y^{+}= {yu_{\tau}}/{\nu}$ where $u_{\tau}$ is so called
friction velocity, y is the absolute distance from the wall, and $\nu$
is the kinematic viscosity. One can interpret $y^+$ as a local Reynolds
number. The friction velocity $u_{\tau}$ is defined as 
\begin{equation}\label{eq:utau}
u_{\tau}=\sqrt{ {\tau_{w}}/{\rho}},
\end{equation}

\noindent where wall shear stress $\tau_w$, $\tau_{w}=\rho\nu\frac{dU}{dy}$
at y=0, and the dimensionless velocity is given by $U^{+}= {u}/{u_{\tau}}$.
The Figure \ref{fig:exp0} demonstrates that 
the superposition $U_{GHE}$ provides an excellent fit to the experimental mean velocity profile for $\gamma$=0.65
and $\delta$=0.052, \citet{Fedoseyev_2023}. 
As to parameter $\gamma$ in Eq.(\ref{eq:GHE1sol}), the method proposed in \citet{Fedoseyev_2023} to obtain $\gamma$ from the momentum equation was approximate. 
In this study, a different approach is proposed, to obtain the parameter $\gamma$ using the minimum dissipation principle for viscous flow. 

The contents of the paper is the following. Section \ref{sec:GHE} presents the GHE and its simplified form for turbulent flow in channel to which the minimization principle to be applied. Section \ref{seq:laminar} shows  a general analytical solution of the GHE, and Section {\ref{sec:principle} (i) formulates a minimization principle, that was chosen from several candidates, (ii) applies the principle to several problems, and (iii) compares the obtained analytical solution with the experimental data for turbulent flows. Section \ref{sec:discussion} provides discussion of the obtained results, which is summarized by the  Conclusions.

%
%

\section{\label{sec:GHE}Governing Equation} 

To proceed with the minimization principle, we have to present the governing equation used, the
Generalized Hydrodynamic Equations (GHE). The GHE are obtained from 
Generalized Boltzmann Transport Equation, \citet{Alexeev_2004},
by multiplying the latter by the standard collision invariants (mass, momentum, and energy), and integrating the result in the velocity space. The particles of finite size are considered.   The obtained equations below are valid for incompressible viscous flow, and have the following non-dimensional form, \citet{Fedoseyev_2012}:

%
%

\begin{equation}
\it \frac{\partial \bf V} {\partial \rm t} + ({\bf V}\nabla) {\bf V}
- Re^{-1}\nabla^2 \bf V + \nabla \rm p - {\bf F} = 
\tau \left\{ 2 \frac{\partial}{\partial t}(\nabla \rm p) + 
\nabla^2 (\rm p \bf V) + \nabla(\nabla \cdot (\rm p \bf V))  \right\},
\label{momeq}
\end{equation}

\begin{equation}
\it \nabla \cdot \bf V = 
\tau \left\{
2 \frac{\partial}{\partial t}(\nabla \cdot {\bf V})
+ \nabla \cdot ({\bf V} \nabla){\bf V}
+\nabla^2 \rm p -\nabla \cdot {\bf F} \right\},
\label{newconteq}
\end{equation}

\noindent
where ${\bf V}$ and $p$ are nondimensional velocity and pressure respectively, ${Re=U_0 L_0/\nu}$ - the Reynolds number, $U_0$ - velocity scale, $L_0$ - 
hydrodynamic length scale, $\nu$ - kinematic viscosity, ${\bf F}$ is
 nondimensional body force and nondimensional timescale ${\tau = \tau^* L_0^{-1}U_0}$. Terms containing $\tau$ are called the fluctuations (temporal and spatial) by \citet{Alexeev_1994}.
One can see the equations become the Navier-Stokes equations if $\tau=0$.

The following assumptions were made by deriving Eq. (\ref{momeq}, \ref{newconteq}):
\begin{enumerate}[label=[(i)]
\item $\tau$ is assumed to be constant.
\item The nonlinear terms of the third order in the
  fluctuations, and the terms of order $\tau$/Re, are neglected. The focus is on large $Re$ numbers.
\item  Assumed slow flow variation, so  second derivatives in time are neglected.
\end{enumerate}

Additional boundary condition was set for pressure on  walls:
\begin{equation}
 {(\nabla \rm p - {\bf F})\cdot {\bf n}= 0} ,
\end{equation}
where ${\bf n}$ is a wall normal.

The GHE  is not a turbulence model, and no additional 
equations are introduced. The solution time of the GHE on a computer is the same as that of the Navier-Stokes equations. By setting $\tau =0$ in the GHE, one actually solves the  Navier-Stokes equations.

%
%

\subsection{\label{sec:2D}GHE for 2D Stationary Incompressible Flow}
The case of 2D incompressible fluid flow in channel is considered  with  the  flow direction in $x$.
For the stationary analytical solution in \citet{Fedoseyev_2023}, GHE (\ref{momeq}), (\ref{newconteq}) were simplified by: 
(a) dropping all temporal derivatives, (b) dropping all the terms (with coefficient $\tau$) in the momentum equations, (c) the nonlinear terms were neglected in the 
fluctuations, (d) all the derivatives in $x$ were neglected, except for the pressure gradient $p_x$=const, so the Laplacian of pressure was $\nabla^{2}p=p_{yy}$.

The resulting continuity equation of GHE model is as follows:
\begin{eqnarray}
v_{y}&=&\tau p_{yy} \label{cont}
\end{eqnarray}

while the momentum equations are :

\begin{eqnarray}
v\,u_{y}+p_{x}&=&Re^{-1} u_{yy} \label{mom1}\\
v\,v_{y}+p_{y}&=&Re^{-1}v_{yy} \label{mom2}
\end{eqnarray}

\noindent where $Re$  is Reynolds number, $\tau$ is given by Eq.\eqref{eq:tau}.
The  boundary conditions are as follows: 
$u=0$, $v=0$ and the normal derivative of pressure $p_n=0$ at the wall $y=0$; $u=U_0$, and $v=0$, $\frac {\partial p}{\partial y} = 0$ (the symmetry conditions) at $y=L/2$.
A symmetry about the centerline $y=L/2$ is assumed, and  the problem is solved in half of the domain.

%
%

\section{General Analytical Solution of GHE for Turbulent Channel Flow\label{seq:laminar}}

It was shown in \citet{Fedoseyev_2023} that the stationary channel flow problem Eq. \eqref{cont},\eqref{mom1},\eqref{mom2} has two solutions. The first is the parabolic velocity $u(y)$ for laminar flow profile :
\begin{eqnarray}
U_{L}=4U_{0}y(L-y)/L^{2}, \label{poiseille}
\end{eqnarray}

\noindent
where $U_{0}=-\frac{1}{8}Re\ p_{x} L^2$ and $v=0$ everywhere.
This solution is
also a solution of the Navier-Stokes equations, as $\nabla^{2}p=0$ in Eq.(\ref{cont}).

However, equations \eqref{cont},\eqref{mom1}, and \eqref{mom2} have a second solution $u(y)$,  which was called a turbulent solution. It was obtained analytically by \citet{Fedoseyev_2023}, and it is a super exponential function:
\begin{eqnarray} \label{eq:UT}
U_{T}=U_{0}\left(1-e^{1-e^{y/\delta}}\right),
\end{eqnarray}

\noindent  and $v(y)=\frac{1}{\delta Re}(1-e^{y/\delta})$, where the parameter $\delta$ is defined by Eq.(\ref{eq:delta}) above.
The Navier-Stokes equations do not have such a solution.

Figure \ref{fig:utul} shows examples of both  $U_L$ (Eq.(\ref{poiseille}), parabolic, green line) and $U_T$ (Eq.(\ref{eq:UT}), super exponential, blue line) solutions for laminar and turbulent flows respectively, and
$U_{GHE}$ solutions for $\gamma=0.6$ (red) and $\gamma=1.2$ (pink). It is interesting to note that $U_T$ for $\delta=0.280$ (red dashed line) coincides  well with the parabolic solution $U_L$ (green).

\begin{figure}
\begin{center}
\includegraphics[width=0.48\textwidth]{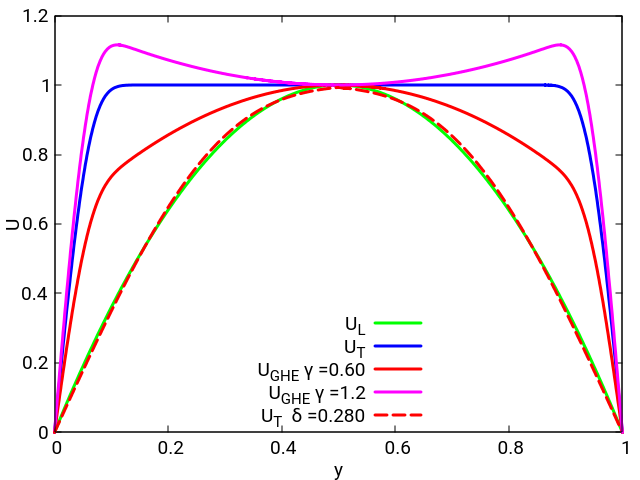}
\end{center}
\caption{\label{fig:utul}Examples of laminar and turbulent solutions, that is $U_L$,
 Eq.(\ref{poiseille}) (parabolic solution, green line), and $U_T$, Eq.(\ref{eq:UT}) (super exponential solution, blue line, $\delta=0.060$) and their superposition $U_{GHE}$ for different $\gamma$. It is interesting to note that $U_T$ for $\delta=0.280$ (red dashed line) coincides well with the parabolic solution.  }
\end{figure}

The general solution for turbulent flow is proposed as a linear superposition of laminar and turbulent solutions, as shown in Eq.\eqref{eq:GHE2sol}:
\begin{equation*}
U_{GHE}=U_{0}\left[\gamma\left(1-e^{1-e^{y/\delta}}\right)+(1-\gamma)4y(L-y)/L^{2}\right]. 
\end{equation*}

The governing equations are nonlinear, and the linear combination above need to be justified. The first term in  brackets, $\gamma U_{T}$,  grows superexponentially in the boundary layer, whereas the second term containing $(1-\gamma) U_L$ is nearly zero, Figure \ref{fig:exp1}. Outside the boundary layer, the first term is constant and the second term starts to grow. 
\begin{figure} 
\begin{center}
\includegraphics[width=0.70\textwidth]{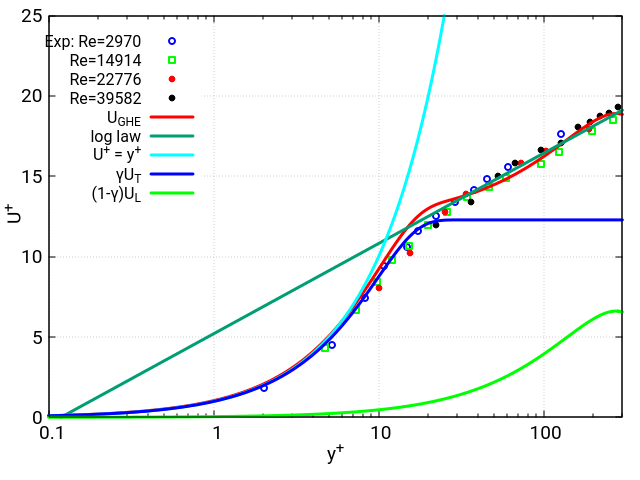}
\end{center}
\caption{\label{fig:law}
Velocity profiles for turbulent flow in channel: experimental data (dots) by \cite{Wei_1989}, analytical solution $U_{GHE}$ (red line) and its constituents, the turbulent solution $\gamma U_{T}$ and the laminar solution $(1-\gamma)U_{L}$; log law by von Karman $U^+$ = 1/k log $y^+$ + B, k = 0.41, B = 5.2 (log law line), and linear law $U^+ = y^+$ (cyan line). }
\label{fig:exp1}
\end{figure}

Therefore  $U_{GHE}$ becomes a function that is approximated by $\gamma U_T$ in the boundary layer and by $\gamma \cdot$ const + $(1-\gamma)U_L$ outside the boundary layer. As in Eq. (\ref{mom1}), (\ref{mom2}) and (\ref{cont}) the function $u=U_{GHE}$ enters only as a derivative in $y$ (all derivatives in $x$ are zero), and the constant disappears. As a result, the derivatives of $u$ are approximated by the derivatives of $\gamma U_T$ in the boundary layer and by the derivatives of $(1-\gamma)U_L$ outside the boundary layer, and the proposed linear superposition of solutions is valid. 

The minimization principle will be applied to the analytical solution $U_{GHE}$ for turbulent flow.
%
%
\section{Minimization Principles for Viscous Fluid Dynamics\label{sec:principle}}

Several papers provided minimization principles for viscous fluid flow: principles of minimum pressure gradient (PMPG), \citet{Taha_2023}, minimum kinetic energy dissipation, \citet{Lyulka_2001},  the principle of minimal viscous dissipation, \citet{Ruang_2022}, and Helmholtz-Korteweg minimization principle, \citet{Borisov_1998}, confirming their ideas by the respective examples. \citet{Lyulka_2001} considered the case of cylindrical pipe with obstacles inside, \citet{Taha_2023} examples included the  unsteady laminar flow in a channel and the flow from harmonically oscillating plate, \citet{Ruang_2022} considered the pressure driven Stokes flow in channel of different cross-sections to confirm  the proposed principles. \citet{Talon_2021} considered the minimization of dissipation as a general principle in physics, and used it for the flow of inelastic non-Newtonian fluids in
macroscopic heterogeneous porous medium. \citet{Borisov_1998} stated that the Helmholtz-Korteweg minimization principle is widely known principle of viscous incompressible fluid mechanics,  analyzed the kinetic energy dissipation functional for internal flows, and found that the stationary point corresponds to the Stokes equations. 

The principle of minimal viscous dissipation was selected in this study.
The dissipation function of a Newtonian fluid with viscosity $\mu$ in 2D channel is
\begin{equation} \label{eq:diss0}
E = 2\mu \left[ \left( \frac{\partial U}{\partial x}\right) ^2 + \left( \frac{\partial V}{\partial y}\right) ^2 -  \frac{ 1}{3} (\nabla \cdot{\bf V})^2 \right] + \mu \left[  \frac{\partial V}{\partial x}  + \frac{\partial U}{\partial y}\right]^2 , 
\end{equation}

\begin{equation}
\nabla \cdot {\bf V} = \frac{\partial U}{\partial x}  + \frac{\partial V}{\partial y}
\end{equation}
\noindent
where the $U$ and $V$ are the components of the dimensional velocity vector,  ${\bf V}$ is the velocity vector, and $x,y$ are streamwise and transversal coordinates.

In the case of a stationary flow in channel all derivatives with respect to $x$ are zero, and the term $\frac{\partial V}{\partial y}$ is small and can be neglected. Therefore, the remaining term for  dissipation is the last term in Eq.\eqref{eq:diss0}, and in our notations, it is as follows:
\begin{equation}\label{eq:diss}
\varepsilon (y) =  \left( \frac{ u_{y}(y)}{u_y(0)}\right)^2,
\end{equation}

\noindent
where the term is made nondimensional using $u_{y}(0)$, the velocity derivative on the wall, in the denominator.
%
%
\subsection{Obtaining parameter $\gamma$ by minimization principle }
Parameter $\gamma$ is obtained using the principle of minimal total viscous dissipation. 
To obtain the total viscous dissipation, one needs to identify the volume to integrate Eq.\eqref{eq:diss}. The chosen volume  is formed by a channel cross-section line, channel walls, and the line where cross-section line will be  in a unit of time, at a distance of the average velocity of the fluid. Therefore, the integral of Eq.\eqref{eq:diss} across the channel is multiplied by the  distance (average velocity, an integral of the velocity  divided by $L$). Total energy dissipation per unit time is expressed as:
\begin{equation}\label{eq:diss_tot}
\varepsilon_{T} =  \frac{ 1}{u_{y}(0)^2} \int^L_0 u_y^2 dy \cdot
\frac{ 1}{L} \int^L_0 u\ dy,
\end{equation}

\noindent
where $u(y)=U_{GHE}$ that depend on $\gamma$, and is given by Eq. \eqref{eq:GHE2sol}. A similar equation was derived in \citet{Horne_1986} (p.6, Eq.(15)).

The dependence of total viscous dissipation on $\gamma$ calculated for the \citet{Wei_1989} experiment is shown in Figure \ref{fig:gamma}(a) at $\delta=0.052$. 
The parameter $\delta$ is related to the material properties $\tau^*$, $\nu$  for distilled water as working fluid,  Eq.\eqref{eq:delta}.

The  dissipation has minimum at $\gamma=0.70$. This is close to the experimental fit $\gamma=0.65$ in \citet{Fedoseyev_2023}. 
The dissipation plot  demonstrates that the laminar flow ($\gamma=0$) has significantly larger dissipation than the turbulent flow ($\gamma > 0$), and  the minimum is almost flat in a range of $\gamma = [0.3,1.3]$. 
%
%
\subsection{Comparison with experiments}

Figure \ref{fig:gamma}(b) shows $\gamma$ calculated for the \citet{Doorne_2007} experiment, with $Re=7200$ and  $\delta=0.047$.  The minimum of dissipation is at $\gamma=0.68$, while the best experimental fit is $\gamma=0.65$ in \citet{Fedoseyev_2023}. 

Figure \ref{fig:doorne}(a) shows the experimental data digitized from \citet{Doorne_2007}, along with several plots: (i) laminar (parabolic) flow profile (green line), (ii) turbulent (superexponential) solution (blue line) 
and (iii) GHE analytical solution (red line). The left part of the GHE plot is for $\gamma=0.68$ (minimal dissipation), and the right part is for $\gamma=0.65$ from \citet{Fedoseyev_2023}. The figure demonstrates that neither
the laminar nor turbulent solution fit the data, but the superposition $U_{GHE}$ provides a good comparison to the experimental data.

Figure \ref{fig:doorne}(b) shows the streamwise velocity in a turbulent channel experiment by \citet{Pasch_2023}, with $Re=14000$, along with the  analytical solution $U_{GHE}$ with coefficient $\gamma=0.62$ obtained by the minimum of  viscous dissipation  principle. Here, $\delta=0.033$, the working fluid is air.

\begin{figure}
\begin{center}
\includegraphics[width=0.48\textwidth]{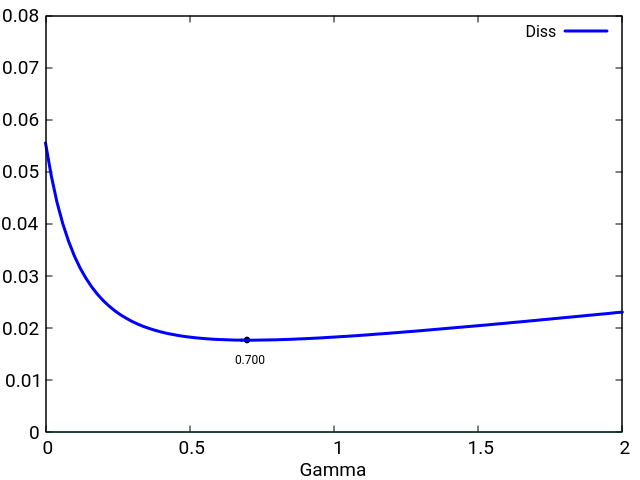}
\includegraphics[width=0.48\textwidth]{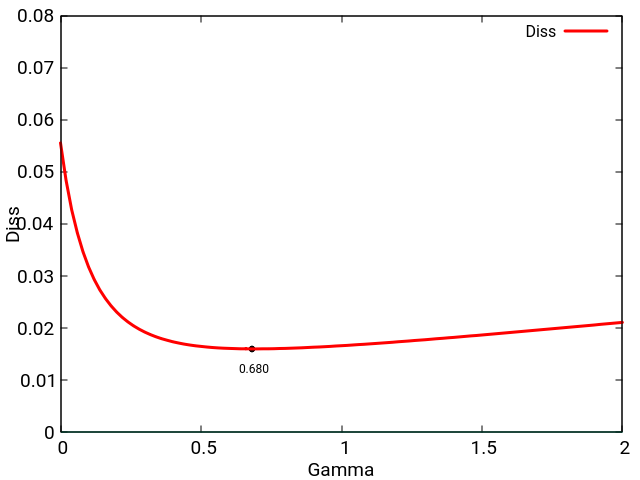}
\end{center}
\hspace{3.5cm} (a) \hspace{6cm}(b) 
\caption{\label{fig:gamma}
(a) Total  viscous dissipation for the \citet{Wei_1989} experiment with $Re=39582$ versus $\gamma$, where $\gamma$ has an extended range [0,2]. The  dissipation has a minimum at  $\gamma=0.70$. Laminar flow ($\gamma=0$) exhibits significantly larger dissipation than turbulent flow. (b) Total  viscous dissipation for \citet{Doorne_2007} experiment in circular pipe, $Re=7200$, the minimum is at $\gamma=0.68$. }
\end{figure}

\begin{figure}
\begin{center}
\includegraphics[width=0.48\textwidth]{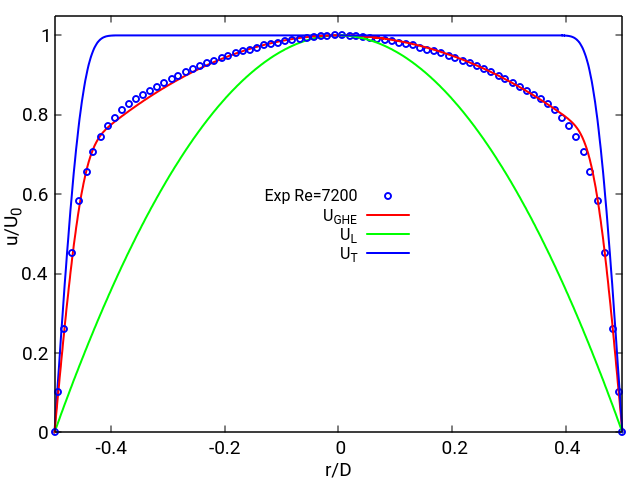}
\includegraphics[width=0.48\textwidth]{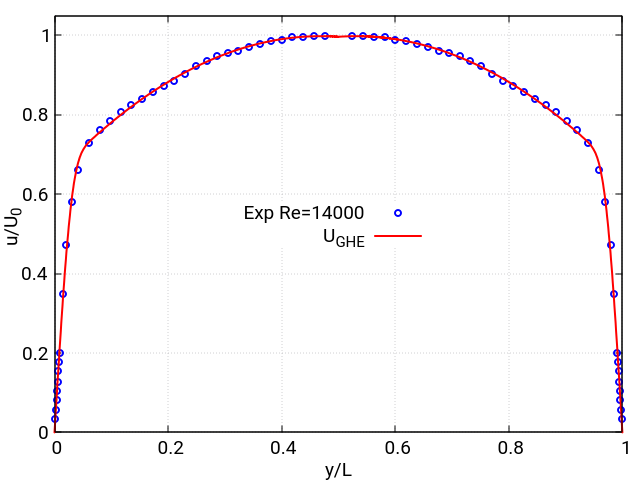}
\hspace{2.5cm} (a) \hspace{6.5cm}(b)
\end{center}
\caption{\label{fig:doorne}
 \textit{(a)} A comparison of the experimental data for the streamwise velocity $U=u/U_0$ versus radius $r/D$  ($D$-diameter) by  \citet{Doorne_2007}, (blue dots) at $Re=7200$, with the GHE solution (red line). The  right part of the $U_{GHE}$ plot is for $\gamma=0.65$ from \citet{Fedoseyev_2023}, and the left part is for $\gamma=0.68$, obtained from the minimum viscous dissipation principle. Both GHE analytical solutions fit the experimental velocity profile well. Also shown are the laminar solution (green line) and the turbulent solution (blue line). \textit{(b)} Comparison of the streamwise velocity in the turbulent channel experiment by \citet{Pasch_2023}, $Re=14000$, (dots) and the analytical solution $U_{GHE}$ with coefficient $\gamma=0.62$ (red line) found by the minimization principle.}
\end{figure}
%
%
\section{Discussion\label{sec:discussion}}
It was shown that the parameter  $\gamma$ can be obtained using the minimization principle. Comparison with several experiments have been provided demonstrating good agreement of the analytical solution with the experimental data.
The GHE solution for turbulent flows depends on two similarity parameters: the Reynolds number $Re$, and the parameter $\delta = {\sqrt{\tau^{*}\nu}} / {L_0}$, which is related to the material properties $\tau^*$ and $\nu$.

%
%
\subsection{Similarity Parameter $\delta$\label{sec:delta}}
The similarity parameter $\delta$ does not depend on the Reynolds number as shown in the case of Wei \& Willmarth experiment, \citet{Wei_1989}, where the experimental data for different Reynolds numbers are falling into the same curve (Figure \ref{fig:exp1}, red line $U_{GHE}$) defined by the parameter $\delta$. 
Knowing $\delta$ one can find the timescale coefficient $\tau^*$, which is the material property.
%
%
\subsection{Timescale Coefficient $\tau$ \label{sec:tau}}
Analyzing several experiments and simulations,  \citet{Wei_1989}, \citet{Koseff_1984}, \citet{Doorne_2007}, and \citet{Fedoseyev_2010}, \citet{Fedoseyev_2012}, \citet{Fedoseyev_2023} we have found that the dimensional timescale coefficient $\tau^* = \delta^2 L_0^2 / \nu$  for distilled water is $\tau^* = 0.40 \pm 0.05$ s, and for tap water is $\tau^* \approx  0.80$ s. There is no estimation of  $\tau^*$ for air, as the viscosity, temperature and pressure were not provided in \citet{Pasch_2023}.
%
%
\subsection{Turbulent Boundary Layer\label{sec:BL}}
One can see that the analytical solution of GHE presents well the turbulent boundary layer in all the regions. 
The linear law is in the range $0 < y^+ < 5$, where the parabolic profile $(1-\gamma)U_{L}$ (laminar solution) is very small, and the analytical solution for small $y/\delta$ becomes 
\begin{eqnarray*}
U_{GHE}&=& \gamma U_{T} = U_{0}\gamma\left(1-e^{1-e^{y/\delta}}\right) =
 U_{0}\gamma y/\delta , 
\end{eqnarray*}
that is a linear law.

The near-middle (buffer) boundary layer region is the range $5 < y^+ < 30$, a strictly nonlinear region, and the analytical solution fits the experiment quite satisfactory.
In the fad-middle (inner) boundary layer region, the range of $30 < y^+ < 200$, the  superexponential part of $\gamma U_{T}$ becomes nearly constant $\approx$ 12.5  (Figure~\ref{fig:law}, blue line), and the GHE analytical solution (red line) changes due to the growth of the laminar part of solution $(1-\gamma)U_{L}$ (Figure~\ref{fig:law}, cyan line). In this region the GHE solution is  $U_{GHE}$ = (12.5 + $(1-\gamma)U_{L}$) and fits well to logarithmic von Karman law, and the experimental data. The outer (non-linear, essentially inviscid) region starts at $y^+ > 200$ and continues to the center line, where the analytical solution fit the experiment well too.

%
%
\section*{Conclusions} 

The method for calculating the coefficients of an approximate analytical solution for turbulent flow in a channel has been presented. These coefficients are obtained using the principle of minimum of viscous dissipation.
The analytical solution was compared with experimental data from several turbulent flow experiments, demonstrating a good agreement with the mean experimental velocity data. 
The obtained analytical solution successfully captured the correct  velocity behavior across the entire turbulent boundary layer and into the external flow,  spanning from the inner viscous sublayer to the outer layer of the boundary layer. 

%
%
\bibliographystyle{jfm}

\end{document}